\documentclass[apj]{emulateapj}
\usepackage{natbib,graphicx,longtable,afterpage}

\shorttitle{Star formation in Type 2 AGN host galaxies}
\shortauthors{Suh et al.}

\begin{document}

\title{Type 2 AGN host galaxies in the {\it Chandra}-COSMOS Legacy Survey: No Evidence of AGN-driven Quenching}

\author{Hyewon Suh\altaffilmark{1}, Francesca Civano\altaffilmark{2}, G\"unther Hasinger\altaffilmark{1}, Elisabeta Lusso\altaffilmark{3,17}, Giorgio Lanzuisi\altaffilmark{4,5}, Stefano Marchesi\altaffilmark{6}, Benny Trakhtenbrot\altaffilmark{7}, Viola Allevato\altaffilmark{8,9}, Nico Cappelluti\altaffilmark{10}, Peter L. Capak\altaffilmark{11,12}, Martin Elvis\altaffilmark{2}, Richard E. Griffiths\altaffilmark{13}, Clotilde Laigle\altaffilmark{14}, Paulina Lira\altaffilmark{15}, Laurie Riguccini\altaffilmark{16}, David J. Rosario\altaffilmark{17}, Mara Salvato\altaffilmark{18}, Kevin Schawinski\altaffilmark{7}, Cristian Vignali\altaffilmark{3,4}}

\altaffiltext{1}{Institute for Astronomy, University of Hawaii, 2680 Woodlawn Drive, Honolulu, HI 96822, USA}
\altaffiltext{2}{Harvard-Smithsonian Center for Astrophysics, Cambridge, MA 02138, USA}
\altaffiltext{3}{INAF Osservatorio Astrofisico di Arcetri, I-50125 Florence, Italy}
\altaffiltext{4}{Dipartimento di Fisica e Astronomia, Universit\`{a} di Bologna, viale Berti Pichat 6/2, 40127 Bologna, Italy}
\altaffiltext{5}{INAF Osservatorio Astronomico di Bologna, Via Ranzani 1, 40127, Bologna, Italy}
\altaffiltext{6}{Department of Physics \& Astronomy, Clemson University, Clemson, SC 29634, USA}
\altaffiltext{7}{Institute for Astronomy, Department of Physics, ETH Zurich, Wolfgang-Pauli-Strasse 27, CH-8093 Zurich, Switzerland}
\altaffiltext{8}{Department of Physics, University of Helsinki, Gustaf H\"allstr\"omin katu 2a, FI-00014 Helsinki, Finland}
\altaffiltext{9}{University of Maryland, Baltimore Country, 1000 Hilltop Circle, Baltimore, MD 21250, USA}
\altaffiltext{10}{Yale Center for Astronomy and Astrophysics, 260 Whitney Avenue, New Haven, CT 06520, USA}
\altaffiltext{11}{Infrared Processing and Analysis Center (IPAC), 1200 E. California Blvd., Pasadena, CA, 91125, USA}
\altaffiltext{12}{California Institute of Technology, 1200 E. California Blvd., Pasadena, CA, 91125, USA}
\altaffiltext{13}{Department of Physics \& Astronomy, University of Hawaii at Hilo, 200 W. Kawili Street, Hilo, HI 96720, USA}
\altaffiltext{14}{Sorbonne Universit\'{e}s, UPMC Universit\'{e} Paris 6 et CNRS, UMR 7095, Institut d'Astrophysique de Paris, 98 bis Boulevard Arago, 75014 Paris, France}
\altaffiltext{15}{Departmanento de Astronom\'{i}a, Universidad de Chile, Casilla 36-D, Santiago, Chile}
\altaffiltext{16}{Observat\'{o}rio do Valongo, Universidade Federal do Rio de Janeiro, Ladeira do Pedro Antonio 43, Sa\'{u}de, Rio de Janeiro, RJ 20080-090, Brazil}
\altaffiltext{17}{Centre for Extragalactic Astronomy, Department of Physics, Durham University, South Road, Durham, DH1 3LE, UK}
\altaffiltext{18}{Max-Planck-Institute f\"ur Plasma Physics, Boltzmann Strasse 2, D-85748 Garching, Germany}


\begin{abstract}
We investigate the star formation properties of a large sample of $\sim$2300 X-ray-selected Type 2 Active Galactic Nuclei (AGNs) host galaxies out to $z\sim3$ in the {\it Chandra} COSMOS Legacy Survey in order to understand the connection between the star formation and nuclear activity. Making use of the existing multi-wavelength photometric data available in the COSMOS field, we perform a multi-component modeling from far-infrared to near-ultraviolet using a nuclear dust torus model, a stellar population model and a starburst model of the spectral energy distributions (SEDs). Through detailed analysis of SEDs, we derive the stellar masses and the star formation rates (SFRs) of Type 2 AGN host galaxies. The stellar mass of our sample is in the range $9 < \log M_{\rm stellar}/M_{\odot} < 12$ with uncertainties of $\sim$0.19 dex. We find that Type 2 AGN host galaxies have, on average, similar SFRs compared to the normal star-forming galaxies with similar M$_{\rm stellar}$ and redshift ranges, suggesting no significant evidence for enhancement or quenching of star formation. This could be interpreted in a scenario, where the relative massive galaxies have already experienced substantial growth at higher redshift ($z>3$), and grow slowly through secular fueling processes hosting moderate-luminosity AGNs. 
\end{abstract}



\section{Introduction}

One of the outstanding issues for understanding the formation and evolution of galaxies is how the presence of a supermassive black hole (SMBH) affects its host galaxy. A connection between the growth of SMBHs and their host galaxies has been widely accepted by observed correlations between black hole mass and host galaxy properties (e.g. \citealt{Magorrian98, Gebhardt00, Ferrarese00, Gultekin09, McConnell13, Kormendy13}), and the remarkable resemblance between the evolutionary behavior of the growth of active galactic nuclei (AGN) and star formation history (e.g. \citealt{Madau96, Giacconi02, Cowie03, Steffen03, Ueda03, Barger05, Hasinger05, Hopkins07, Aird15, Caplar15}). The existence of these correlations seems to support that nuclear activity and star formation might co-exist with the same gas reservoir fueling black hole accretion and star formation simultaneously (e.g. \citealt{Springel05, Netzer09, Mullaney12, Rosario13, Vito14}). However, our current understanding of the effects that AGN can have on the star formation processes is still under debate (see \citealt{Alexander12, Kormendy13, Heckman14} for recent reviews). 

There has been a general consensus that the majority of star-forming galaxies show a tight correlation between the star formation rate (SFR) and their stellar mass, commonly referred to as the main sequence (MS) of star formation (e.g. \citealt{Noeske07, Daddi07, Elbaz07, Rodighiero11, Whitaker12, Rodighiero14}). \citet{Speagle14} present the calibrated relationship between stellar mass and SFR out to $z\sim6$ using a compilation of 25 star-forming MS studies in a variety of fields, reporting that the MS galaxies have a $\sim$0.2 dex scatter in the slope of their M$_{\rm stellar}$-SFR relation and remains constant over cosmic time. The existence and tightness of this star formation sequence can be interpreted assuming that the growth of the majority of star-forming galaxies have been regulated more by internal secular processes rather than by merger process (e.g. \citealt{Elbaz11, Rodighiero11, Wuyts11}). 

Controversial results were found for AGN host galaxies: some studies have indicated equivalent or enhanced star formation compared to normal star-forming galaxies (e.g. \citealt{Silverman09, Mullaney12, Rosario12, Santini12, Juneau13}), whereas some others have shown that AGN host galaxies lie below the MS of star-forming galaxies, suggesting that AGN accretion might suppress and eventually quench star formation via a process of feedback (e.g. \citealt{Barger15, Mullaney15, Riguccini15, Shimizu15}). Furthermore, several studies have addressed that the majority of AGN host galaxies in the local universe are preferentially found in the green valley on the color-magnitude diagram, transitioning from star-forming galaxies in the blue cloud to passive galaxies on the red sequence (e.g. \citealt{Schawinski09, Trump13}). Therefore, the question of whether AGN activity can significantly enhance or quench star formation in galaxies is still unsettled. 

Different results can be produced by either physical properties of the sources, or observational biases, or both. The sample selection including completeness and biases due to a specific selection method (X-ray versus infrared selected AGNs, for example), as well as the use of different SFR indicators could introduce systematics since the contribution of AGN emission may significantly hamper the precise determination of SFRs of AGN host galaxies. The {\it Herschel} Space Observatory \citep{Pilbratt10, Poglitsch10, Griffin10} covers the far-infrared emission from dust including the characteristic far-IR bump typically seen in star-forming galaxies, allowing us for more precise measurements of the total IR luminosity, especially for dusty galaxies and AGN host galaxies, since many of the often used SFR indicators (e.g. H$\alpha$, UV continuum) can be substantially contaminated by AGN-related emission (e.g. \citealt{Dale07, Schweitzer07, Netzer07, Lutz16}). 

Deep, large-area X-ray observations with {\it Chandra} in the COSMOS field (i.e. {\it Chandra}-COSMOS, {\it Chandra}-COSMOS Legacy Survey; \citealt{Elvis09, Civano16}) open up a new regime for studying a large sample of the AGN population over a broad range of luminosities (${\rm 41<log~L_{0.5-10~keV} (erg/s) <45}$) out to $z\sim5$, providing a unique opportunity of studying AGN evolution. X-ray surveys are the most efficient way for selecting AGNs over a wide range of luminosities and redshifts because they are less affected by obscuration, and also the contamination from non-nuclear emission, mainly due to star-formation processes, is far less significant than in optical and infrared surveys \citep{Donley08, Donley12, Lehmer12, Stern12}. Therefore, the X-ray emission is a relatively clean signal from the nuclear component. Furthermore, using the AGN sample with the large, uniform X-ray depth and coherent observations in the COSMOS field, we can minimize the systematic selection effects (e.g. \citealt{Lauer07, Rosario13, Caplar15}). The already existing extensive compilation of multi-wavelength data in the COSMOS field allows us to investigate AGN host galaxies to have a better understanding of nuclear activity and its connection with star formation. 

In this paper, we investigate the properties of AGN host galaxies in the {\it Chandra} COSMOS-Legacy survey. Since the SMBH-powered emission contributes significantly to the ultraviolet-to-optical parts of the spectra of Type 1 AGNs (e.g. \citealt{Elvis12, Hao13}), it is extremely difficult to determine reliable stellar mass for Type 1 AGNs (e.g. \citealt{Maiolino10}). A detailed analysis of the Type 1 AGN host galaxies in the {\it Chandra} COSMOS Legacy Survey, including spectroscopic analysis in the optical and near-infrared wavelength ranges, will be presented in a second paper, Suh et al. (in preparation). Thus, we focus on the non-broad-line and/or obscured AGN host galaxies based either on the spectroscopic or the photometric classification. Hereafter, we refer to non-broad-line and/or obscured sources as ``Type 2" AGNs. We utilize multi-wavelength data from near-ultraviolet to far-infrared of a large sample of AGNs in the {\it Chandra} COSMOS Legacy field. We use the Spectral Energy Distribution (SED) fitting to disentangle the galaxy and nuclear contributions in order to measure the stellar masses and the SFRs accurately. Finally we discuss the effects of the nuclear activity on the star formation in Type 2 AGN host galaxies by comparing to the normal star-forming galaxies. 

Throughout this paper we assume a $\Lambda$CDM cosmology with $\Omega_{m}=0.3,~\Omega_{\Lambda}=0.7$, and $H_{0}=70~{\rm km~s^{-1}~Mpc^{-1}}$.


\section[]{X-ray-selected AGN Sample}

\begin{figure}
\centering
\includegraphics[width=0.5\textwidth]{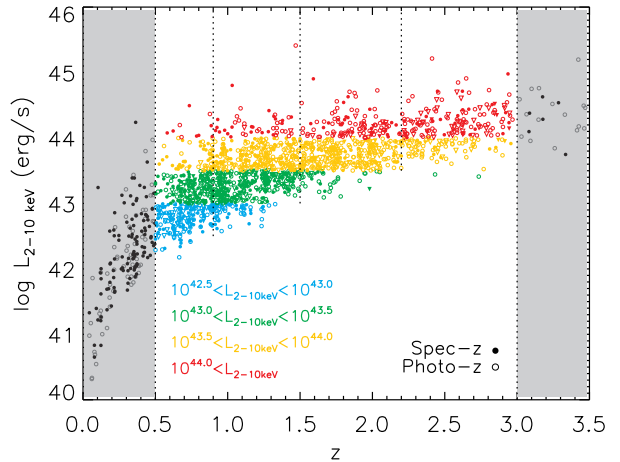}
\caption{The absorption-corrected X-ray (2-10 keV) luminosity versus spectroscopic (solid circle) and/or photometric (open square) redshift for our sample of Type 2 (non-broad-line or obscured) AGNs from CCLS. We split our sample into four redshift bins (vertical dashed lines). Colored symbols indicate sources in different L$_{2-10~keV}$ bins. For sources which are not detected in hard band but in other bands (full and/or soft), upper limits of L$_{2-10~keV}$ are shown with downward triangles.}
\end{figure}

\subsection{The {\it Chandra} COSMOS Legacy Survey}

The {\it Chandra} COSMOS-Legacy Survey (CCLS; \citealt{Civano16}) is a large area, medium-depth X-ray survey covering $\sim$2 deg$^{2}$ of the COSMOS field obtained by combining the 1.8 Ms {\it Chandra} COSMOS survey (C-COSMOS; \citealt{Elvis09}) with 2.8 Ms of new {\it Chandra} ACIS-I observations. The CCLS is wide enough to have one of the largest samples of X-ray AGNs selected from a single contiguous survey region, containing 4016 X-ray point sources, and also deep enough to find faint sources down to limiting fluxes of ${\rm 2.2\times10^{-16}~erg~cm^{-2}~s^{-1}}$, ${\rm 1.5\times10^{-15}~erg~cm^{-2}~s^{-1}}$, ${\rm 8.9\times10^{-16}~erg~cm^{-2}~s^{-1}}$ in the soft (0.5-2 keV), hard (2-10 keV), and full (0.5-10 keV) bands. Moreover, CCLS sources are bright enough so that 97\% of these were identified in the optical and infrared bands and therefore photometric redshifts were computed. Thanks to the intense spectroscopic campaigns in the COSMOS field, $\sim$54\% of the X-ray sources have been spectroscopically identified and classified. The full catalog of CCLS has been presented by \citet{Civano16} and \citet{Marchesi16}, including X-ray and optical/infrared photometric and spectroscopic properties.

We select a sample of Type 2 AGNs using the spectroscopic type when available (sources classified as non-broad-line AGN, which show only narrow emission line and/or absorption line features in their spectra), or the photometric type (sources which are fitted either with an obscured AGN template or with a galaxy template) from the catalog of X-ray point sources in the CCLS \citep{Marchesi16}. 2716 sources are classified as Type 2 AGNs with spectroscopic redshifts (1027) or photometric redshifts (1689). We compute the absorption-corrected X-ray luminosity of Type 2 AGNs using absorption-correction factor from \citet{Marchesi16} which is obtained assuming an X-ray spectral index $\Gamma$=1.8. In Figure 1, we show the absorption-corrected 2-10 KeV X-ray luminosity L$_{2-10~keV}$ of Type 2 AGNs as a function of redshift (spectroscopic or photometric). We estimate L$_{2-10~keV}$ values using upper limits for sources which are not detected in the hard band but detected in the full band. 1980 sources have been detected in the full band (1618 in the soft and 1532 in the hard band). Sources with photometric and spectroscopic redshifts are indicated with open and solid circles, respectively. The final sample analyzed in this paper consists of 2267 out of 2716 Type 2 AGNs in the redshift range $0.5<z<3.0$, in order to avoid effects related to the flux- or volume-limits of the survey, and because at $z>3.0$ sources only have partial spectral coverage which makes it challenging to perform a statistically significant SED fitting. Our Type 2 AGN sample has X-ray luminosities spanning 3 orders of magnitude from $10^{42}$ to $10^{45}$ erg/s in the hard band. Colored symbols indicate sources in different X-ray luminosity (L$_{2-10~keV}$) bins. For sources which are not detected in the hard band but in the full band, we show upper limits of L$_{2-10~keV}$ with downward triangles.

\subsection{Multi-wavelength dataset}

\begin{deluxetable}{lr}
\tabletypesize{\scriptsize}
\tablewidth{0pt}
\tablecaption{Detection Fraction for each Photometry Bands}
\tablehead{
    \colhead{\textbf{Photometry Band}} & 
    \colhead{\textbf{Detection fraction}}}
\startdata
  {\it GALEX} NUV & 3\%(64/2267)   \\
  CFHT U & 62\%(1395/2267) \\
  Subaru B & 72\%(1634/2267) \\
  Subaru V & 72\%(1637/2267) \\
  Subaru r & 82\%(1866/2267) \\
  Subaru i & 84\%(1896/2267) \\
  Subaru z$^{+}$ & 87\%(1964/2267) \\
  UltraVista Y & 72\%(1640/2267) \\
  UltraVista H & 75\%(1697/2267) \\
  UltraVista J & 78\%(1761/2267) \\
  UltraVista Ks & 79\%(1798/2267) \\
  {\it Spitzer} 3.6$\mu$m & 92\%(2079/2267) \\
  {\it Spitzer} 4.5$\mu$m & 92\%(2079/2267) \\
  {\it Spitzer} 5.8$\mu$m & 86\%(1946/2267) \\
  {\it Spitzer} 8.0$\mu$m & 76\%(1721/2267) \\  
   \enddata
\end{deluxetable}

We compile the SEDs of our sample of Type 2 AGNs from near-ultraviolet (near-UV; 2300\AA) to far-infrared (far-IR; 500$\mu$m) wavelengths using the multi-wavelength photometric data available in the COSMOS field. Specifically, we use the most recent photometric catalog from \citet{Laigle16} including the {\it GALEX} near-UV band, CFHT U band, five Subaru Suprime-Cam bands (B, V, r, i, z$^{+}$), four UltraVista bands ({\it Y, H, J, Ks}), and four {\it Spitzer}/IRAC bands (3.6, 4.5, 5.8 and 8.0$\mu$m). The detection fraction for each photometry bands is presented in Table 1. In addition, we use the 24$\mu$m and 70$\mu$m Multiband Imaging Photometer for {\it Spitzer} (MIPS) bands \citep{Sanders07, LeFloch09} with $\sim$59\% (1329/2267) of the sources detected in the 24$\mu$m band, which is particularly important for identifying AGN dusty obscuring structure. We also constrain the SEDs in the far-IR wavelength range for $\sim$25\% (568/2267) of the sources which have been detected by the {\it Herschel} Space Observatory (PACS 100$\mu$m ($\sim$12\%; 262/2267), 160$\mu$m ($\sim$10\%; 222/2267) and SPIRE 250$\mu$m ($\sim$20\%; 451/2267), 350$\mu$m ($\sim$10\%; 224/2267), 500$\mu$m ($\sim$3\%; 60/2267); \citealt{Pilbratt10, Poglitsch10, Griffin10}). We limit the work to only those objects with at least five detected photometric data points ($\sim$91\%; 2056/2267), to guarantee a reliable measurement of the SED fits.

\section[]{Spectral Energy Distribution Fitting}

To derive the physical properties of AGN host galaxies, we developed a 3-component SED fitting technique which allows to disentangle the nuclear emission from the stellar light. Over the far-IR to near-UV wavelength coverage, we decompose the entire SED into a nuclear AGN dusty obscuring structure (i.e., torus), a host galaxy with stellar populations, and a starburst component, which is crucial for estimating reliable physical properties of host galaxies such as galaxy mass and SFR. The method used here is similar to the one applied by \citet{Lusso11} and \citet{Bongiorno12} on the XMM-COSMOS dataset, with significant improvements, including the Bayesian method described in the following sections.

\subsection{Model templates}

\begin{figure}
\centering
\includegraphics[width=0.5\textwidth]{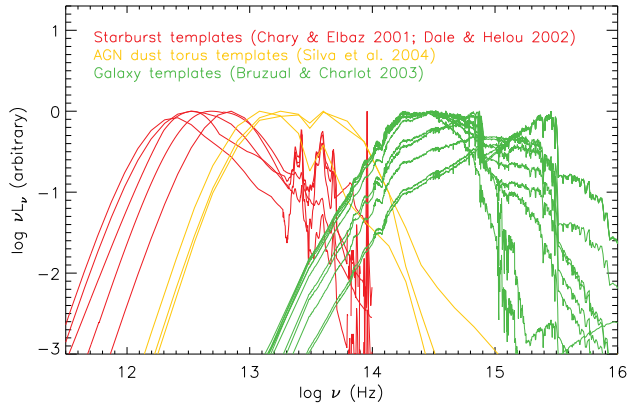}
\caption{Examples of model templates used in the multi-component SED fitting. Green curves indicate some examples of host galaxy templates with various combinations of $\tau$=[0.1, 1, 3], and t$_{\rm age}$=[50 Myr, 2 Gyr] with E(B-V)=[0.0, 0.3]. Yellow curves correspond to three AGN dust torus templates depending on the hydrogen column density, N$_{\rm H}$. Red curves correspond to the subset of starburst templates.}
\end{figure}

In order to examine the SEDs for Type 2 AGN host galaxies, we use model SEDs which are made by combining a stellar population, hot dust emission from AGN (torus) and IR starburst templates to match the broad-band photometry SEDs of our sample. The nuclear emission in obscured AGN dominates the SED only in the X-ray band, while at other wavelengths, the light is mainly due to the galaxy emission combined with reprocessed nuclear emission in the near-IR and mid-IR. While nuclear emission, reprocessed by dust, could significantly contribute to the mid-IR luminosity, the far-IR luminosity is known to be dominated by galaxy emission produced by star formation activity (e.g. \citealt{Kirkpatrick12}). Although a recent study by \citet{Symeonidis17} pointed out that the most powerful unobscured quasars could dominate the far-IR luminosity, we only consider the far-IR luminosity produced by starburst activity for our sample of moderate-luminosity AGNs.

The optical SED of a galaxy represents the integrated light of the stellar populations. We have generated a set of synthetic spectra from the stellar population synthesis models of \citet{BC03}. We have used solar metallicity and the Chabrier initial mass function \citep{Chabrier03}. We have built 10 exponentially decaying star formation histories (SFH), where the optical star formation rate is defined as SFR $\propto~e^{t/\tau}$, with characteristic times ranging from $\tau=0.1$ to 30 Gyr, and a model with constant star formation. For each SFH, the SEDs are generated by models with 15 grids of ages (t$_{\rm age}$) ranging from 0.1 Gyr to 10 Gyr, with the additional constraint on each component that the age should be smaller than the age of the Universe at the redshift of the source. The library of stellar population models is composed by 165 templates. Since the stellar light can be affected by dust extinction, we take into account the reddening effect using the \citet{Calzetti00} law. We have considered E(B-V) values in the range between 0 and 0.5 with steps of 0.05, and the range between 0.5 and 1 with a step of 0.1. We show some examples of stellar population templates with various combinations of $\tau$=[0.1, 1, 3], and t$_{\rm age}$=[50 Myr, 2 Gyr] with E(B-V)=[0.0, 0.3] in Figure 2 (green curves).  

In general, the SED of an obscured AGN is characterized by the near-infrared bump that is a result of the absorption of intrinsic nuclear radiation by dust clouds in the proximity of the central region (so-called torus) on parsec scales, which subsequently re-radiate at infrared frequencies \citep{Barvainis87}. The dust torus SED templates are taken from \citet{Silva04}, as constructed from the study of a large sample of Seyfert galaxies for which clear signatures of non-stellar nuclear emission were detected in the near-IR and mid-IR, and also using the radiative transfer code GRASIL \citep{Silva98}. There are three different templates depending on the amount of nuclear obscuration in terms of hydrogen column density, ${\rm 10^{22} < N_{H} < 10^{23}~cm^{-2}}$, ${\rm 10^{23} < N_{H} < 10^{24}~cm^{-2}}$, ${\rm N_{H} > 10^{24}~cm^{-2}}$ for Seyfert 2. The three templates of Type 2 AGN dust torus are plotted in Figure 2 with yellow curves. The larger the column density, the higher is the nuclear contribution to the IR emission. Although the X-ray data for our AGNs contains some information on the ${\rm N_{H}}$ towards each source (see \citealt{Marchesi16}), we chose to allow ${\rm N_{H}}$ to be a free parameter in our SED fitting.  

For the starburst component in the far/mid-IR region, we adopted 169 starburst templates (105 from \citealt{Chary01} and 64 from \citealt{Dale02}) for fitting the cold dust emission (i.e. far-IR emission). It has been shown that measuring the far-IR luminosity from fitting the far-IR region to libraries of SED \citep{Chary01, Dale02} gives roughly the same results as the modified blackbody plus power-law model \citep{Casey12, U12, Lee13}. The \citet{Chary01} templates are generated based on the SEDs of four prototypical starburst galaxies (Arp220, ULIRG; NGC6090, LIRG; M82, starburst; and M51, normal star-forming galaxy). The \citet{Dale02} templates are based on 69 normal star-forming galaxies, representing a wide range of SED shapes and IR luminosities, complementing each other. A small subset of starburst templates are shown in Figure 2 as red curves.

\begin{figure*}
\centering
\includegraphics[width=0.95\textwidth]{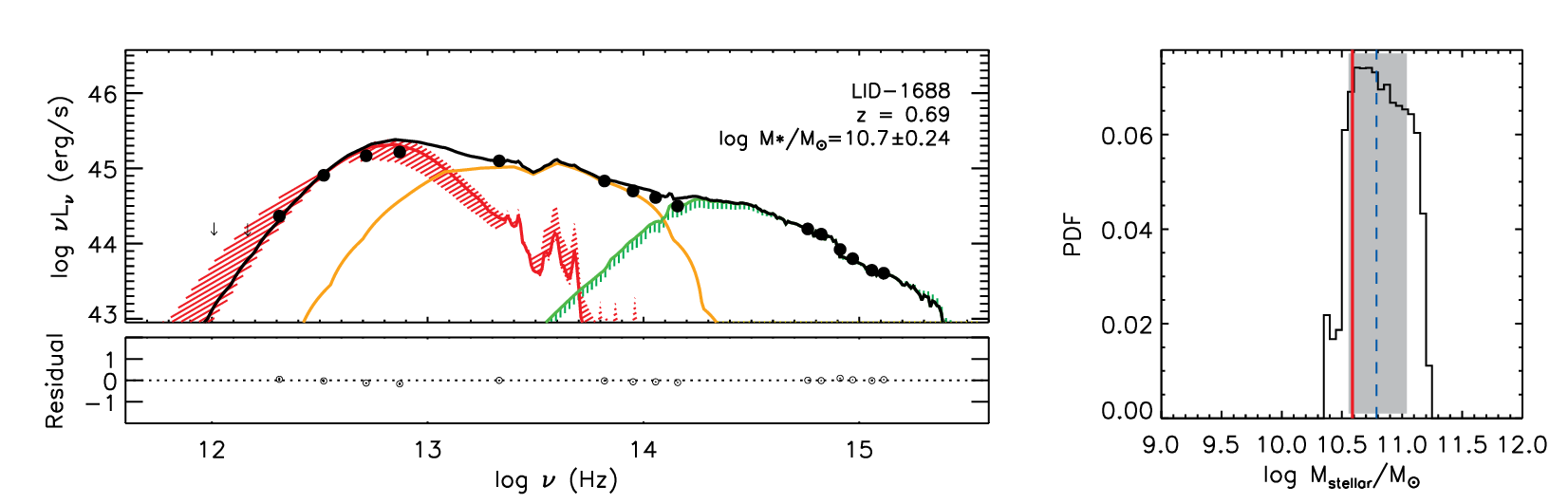}
\includegraphics[width=0.95\textwidth]{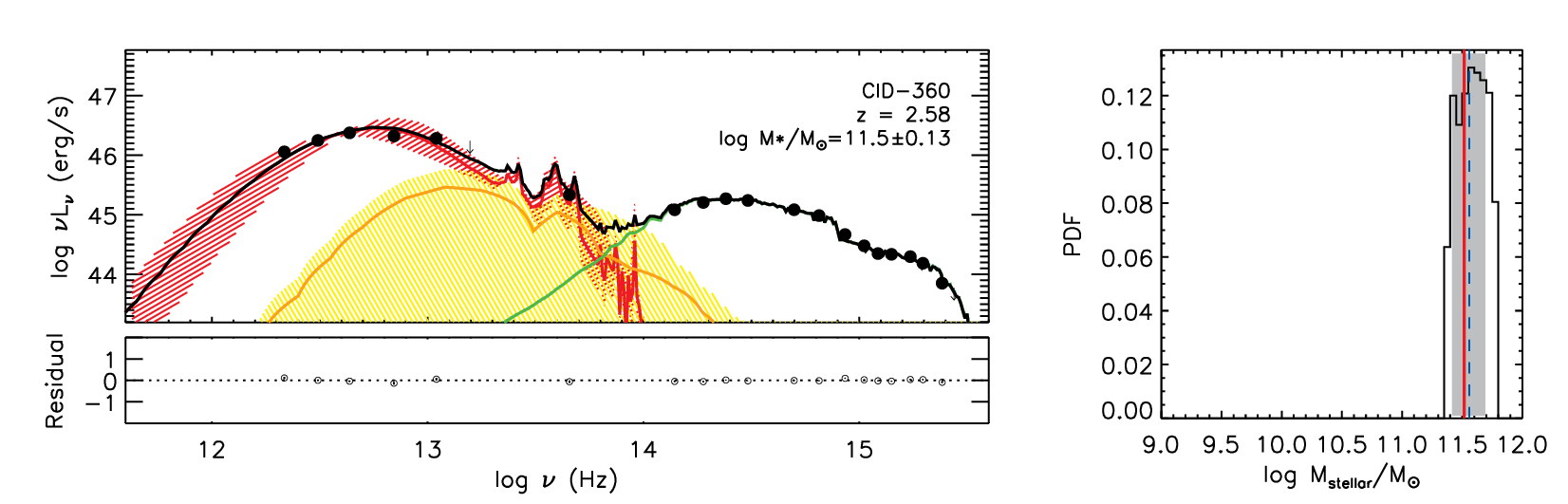}
\includegraphics[width=0.95\textwidth]{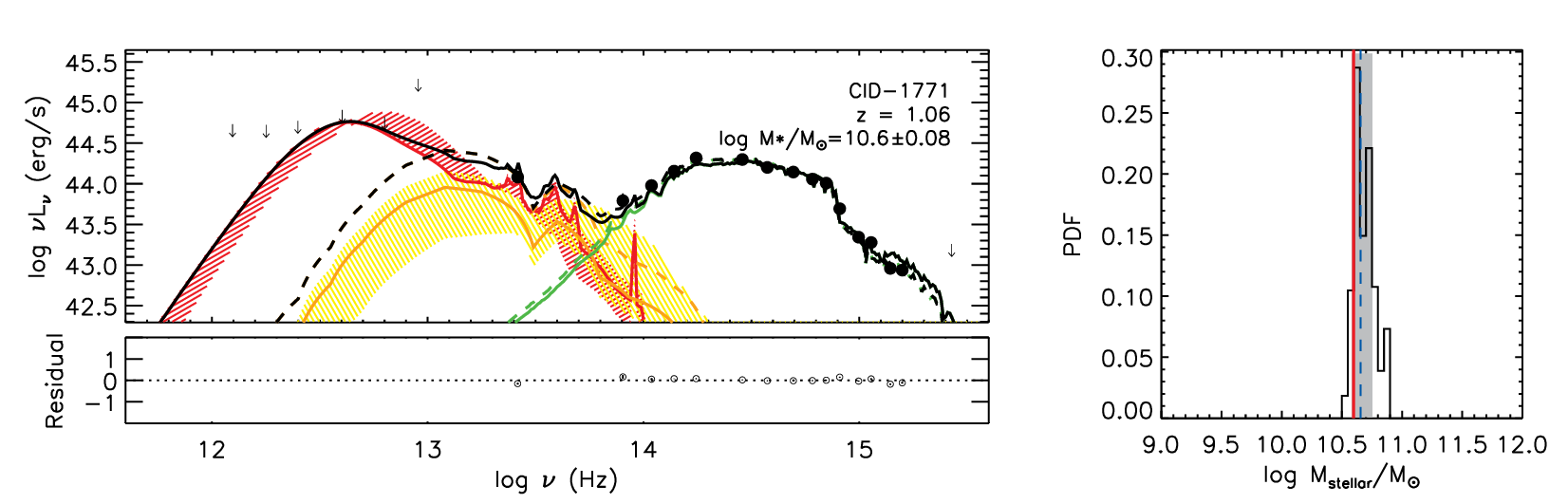}
\includegraphics[width=0.95\textwidth]{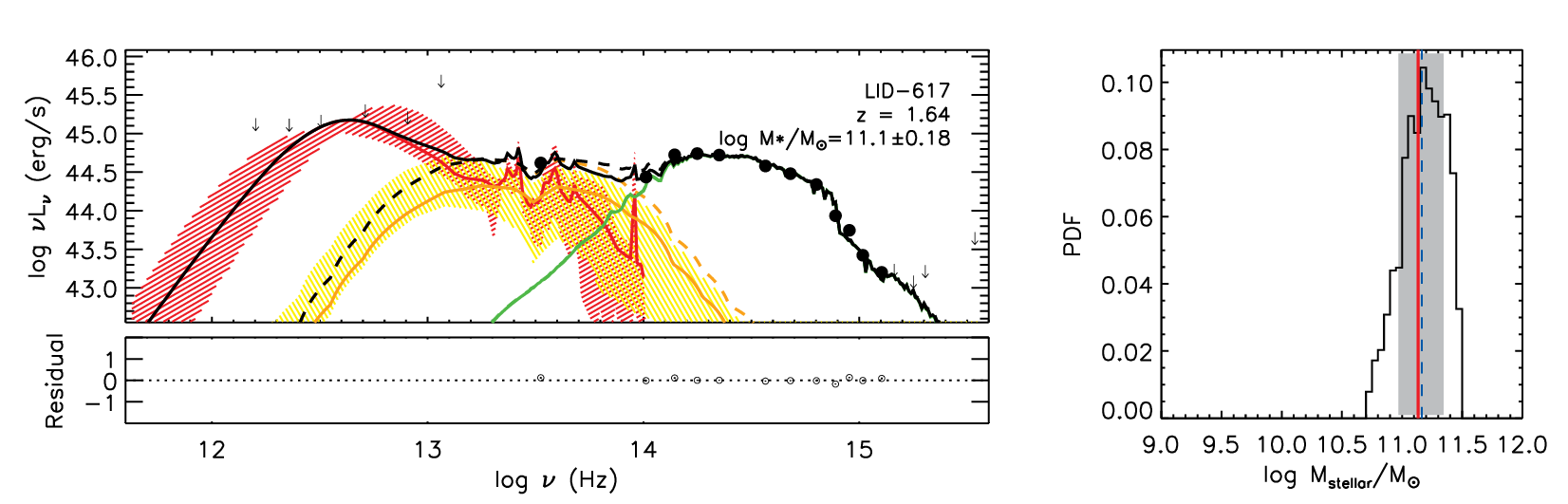}
\caption{Examples of the SED fits (left panels) for sources which are detected in {\it Herschel} far-IR photometry (LID-1688, CID-360), and sources which are detected in 24$\mu$m MIPS photometry but faint in the far-IR (CID-1771, LID-617). The rest-frame observed photometric data (black points) and the detection limits (arrows) are shown with the best-fit model (black solid curve). For the far-IR faint sources (CID-1771, LID-617), we show two different best fit models (solid and dashed curves). The galaxy template (green), AGN dust torus template (yellow), and starburst component (red) are also indicated. The residuals are shown in the lower plot of each spectrum. In the right panels, we show the PDFs for the stellar mass of each source. The best-fitting values are shown in red solid line. The expectation values (blue dashed) and the 16 and 84 percentile intervals (gray shades) are also indicated.
}
\end{figure*}

\subsection{Multi-component SED fitting}

We developed a 3-component SED fitting procedure in which the observed photometric data is fitted at a fixed redshift of the source with a large grid of models obtained by combining the templates described above. The observed flux can be expressed as the sum of 3 components as
\begin{equation}
f_{obs} = C_{1}f_{\rm stellar population} + C_{2}f_{\rm AGN} + C_{3}f_{\rm starburst}
\end{equation}
where the $C_{1}$, $C_{2}$, and $C_{3}$ are coefficients that reproduce the observed data by $\chi^{2}$ minimization. The best-fit SED solution could be a stellar population with a negligible contribution from AGN/starburst components, or a stellar population with the central AGN component, or a stellar population with starburst component, or a stellar population with both AGN and starburst components. 

The fit is performed differently for sources detected in the far-IR and those that are not. Specifically, for the sources detected at 24$\mu$m but not in any far-IR {\it Herschel} wavelength, there are large uncertainties in the estimate of $C_{2}$ and $C_{3}$, because both could substantially contribute in the observed 24$\mu$m band, introducing a degeneracy in the SED fitting. This implies that the fitting can produce two different probable solutions with a similar $\chi^{2}$. One is a prominent AGN dominating in the IR range with no contribution from the dust emission heated by stars, and the other is a negligible AGN contribution in the 24$\mu$m band with the infrared emission dominated by star-forming regions. Therefore, we perform two different fits for the sources which are not detected at any far-IR wavelength. (1) the best-fit model with a possible star-forming component using {\it Herschel} upper limits, adopting the same approach as described by \citet{Calistro16}. Specifically, we consider {\it Herschel} detection limits in each {\it Herschel} band (${\rm flux_{limit}}$) to make mock data points in the far-IR wavelength range, assuming the flux to be ${\rm flux_{limit}}$/2 with an uncertainty $\pm~{\rm flux_{limit}}$/2, to fit the possible star-forming component. (2) We assume a negligible contribution from star formation in the IR range, ${\rm L_{FIR}}$=0, and a significant contribution from the AGN at 24$\mu$m. Thus, we have a range of possible ${\rm L_{FIR}}$ values for {\it Herschel}-undetected sources (i.e., minimum to maximum).

We show examples of the SED fits for the sources which are detected in far-IR photometry (top two panels; LID-1688, CID-360), and the sources which are undetected in the far-IR (bottom two panels; CID-1771, LID-617) in the left panels of Figure 3. The rest-frame photometric data (black points) and the detection limits (arrows) are shown with the best-fit model (black solid curve). For the far-IR faint sources (CID-1771, LID-617), we show two different best fit models in the IR wavelength range: possible star-forming component using upper limits (solid curve) and negligible star formation contributions (dashed curve). The galaxy template (green), the AGN dust torus template (yellow), and the starburst component (red) are also indicated. The residuals are also shown in the lower panel of each SED fit. 

The $\chi^{2}$ minimization is used to determine the best fit among all the possible template combinations. However, its absolute value is not a reliable indicator, because systematic uncertainties may dominate the statistical errors. Therefore, we compute a complementary statistic on the quality of fit, which is the variation of the residual from the fit. We remove $\sim$1\% (27/2056) sources which show large variations in their residuals ($>0.5$), since this indicates that their high $\chi_{red}^{2}$ is not due to an underestimation of the photometric errors but either caused by the lack of suitable templates or by the bad photometry.

\subsection{Estimation of physical parameters}

While the use of the $\chi^{2}$ minimization technique can give an indication of the overall quality of the fitting, the best-fit value could not be a good estimate of representative of physical parameter values in a multi-dimensional parameter space with degeneracies. We, therefore, use Bayesian statistics to derive the most representative value for each parameter of galaxy physical properties, and to evaluate the robust uncertainties since it accounts for the degeneracies inherent in our SED templates.

\begin{figure}
\centering
\includegraphics[width=0.5\textwidth]{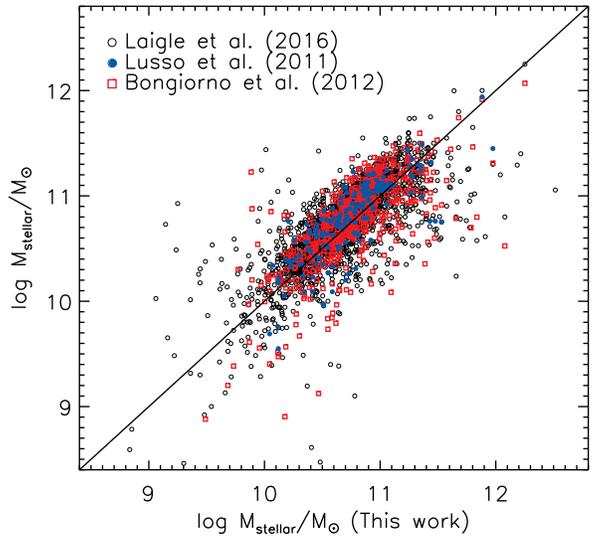}
\caption{Comparison between stellar masses derived from our SED fitting and that from Lusso et al. (2011; blue circles), Bongiorno et al. (2012; red squares), and Laigle et al. (2016, Le Phare; black open circles). The black line denotes a one-to-one relation.}
\end{figure}

We explore any possible combination of SED parameters, which includes the age since the onset of star formation, the e-folding time $\tau$ for exponential SFH models, and the dust reddening. We take into account the possible range for each parameter (i.e. for galaxy mass, $7 < \log\left(M_{\rm stellar}/M_\odot\right) < 13$), and find all the models that produce a value for the parameter. We then build a probability distribution function (PDF) for the stellar mass with the likelihood, $\exp(-0.5~\chi^{2})$, associated with that model for a given source. We estimate expectation values and uncertainties as the width of the parameter values corresponding to the 16 and 84 percentile of the cumulative PDF. In the right panels of Figure 3, we show PDFs for the stellar mass for each of the example sources. In each case, the best-fitting values are shown as red solid line. We also show the expectation values (blue dashed) and the 16 and 84 percentile intervals (gray shades) derived from the cumulated PDFs. We note that the expectation and the best-fitting values are usually very close to each other. In Figure 4, we show the comparison of the stellar masses obtained from our SED fitting with the results from \citet{Lusso11} (blue circles) and \citet{Bongiorno12} (red squares) based on their SED fitting, and Le Phare pipeline products (\citealt{Laigle16}; black open circles). While our sample explores a broader range of redshifts and luminosities, we find good agreements on the stellar masses of matched sources, mainly bright AGNs. The 1$\sigma$ dispersions between the stellar mass derived in this work and other works are 0.27 dex \citep{Lusso11} and 0.30 dex \citep{Bongiorno12}, respectively.

The SFR is estimated by combining the contributions from UV and IR luminosity, which can estimate reliable total SFR since dust in the galaxy is heated by UV emission produced by young massive stars, and then re-emitted in the mid-to-far infrared regime (see e.g. \citealt{Draine03}). We derive the total SFR conversion using the relation from \citet{Arnouts13}, which is similar to that proposed by \citet{Bell05} and adjusted for a \citet{Chabrier03} Initial Mass Function (IMF),
\begin{equation}
{\rm SFR_{total}~(M_{\odot}/yr)} = (8.6\times10^{-11}) \times (L_{IR}/L_{\odot}+2.3 \times \nu L_{\nu}(2300\AA))
\end{equation}
where $L_{IR}$ is the total rest-frame star-forming IR luminosity, which is integrated between 8-1000$\mu$m from the starburst template, and $L_{\nu}$(2300\AA) represents the rest-frame intrinsic absorption-corrected near-UV luminosity at 2300\AA~in units of L$_{\odot}$. To account for {\it Herschel}-undetected sources, we derive upper limits on their SFRs by assuming possible star-forming IR luminosity from the best-fit using {\it Herschel} detection limits. In addition, we also derive the minimum SFRs using only UV luminosity, assuming $L_{IR}=0$. Therefore, we have a range of possible values for SFRs for {\it Herschel}-undetected sources (i.e. from minimum to the maximum SFRs). In Table 2, we present Type 2 AGN host galaxy properties which include stellar masses, SFRs, and luminosities, derived from the SED fitting.

\begin{figure*}
\centering
\includegraphics[width=1\textwidth]{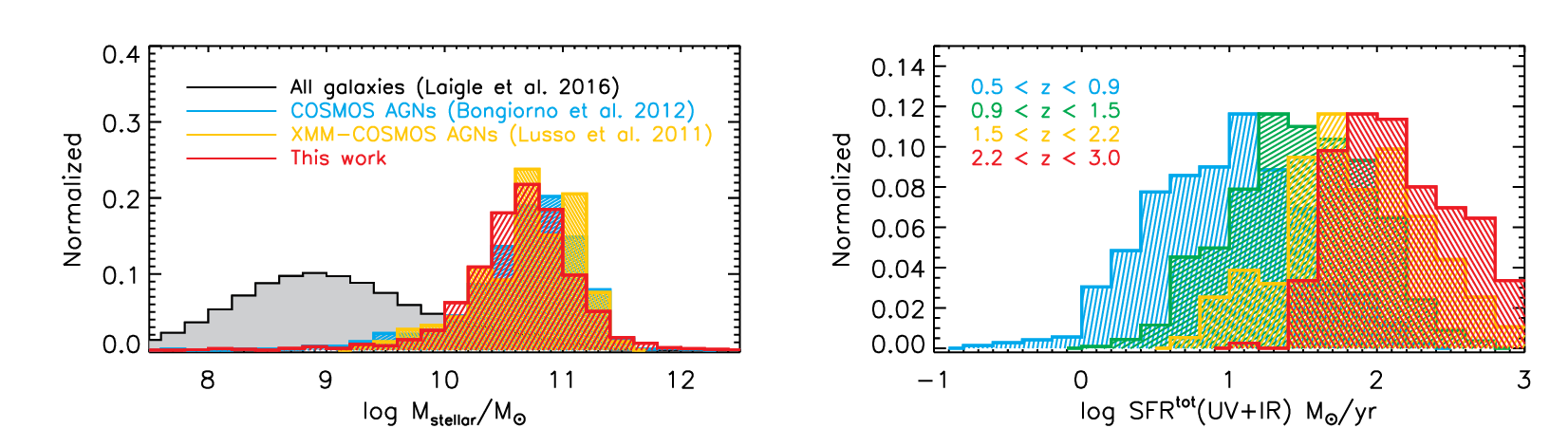}
\includegraphics[width=1\textwidth]{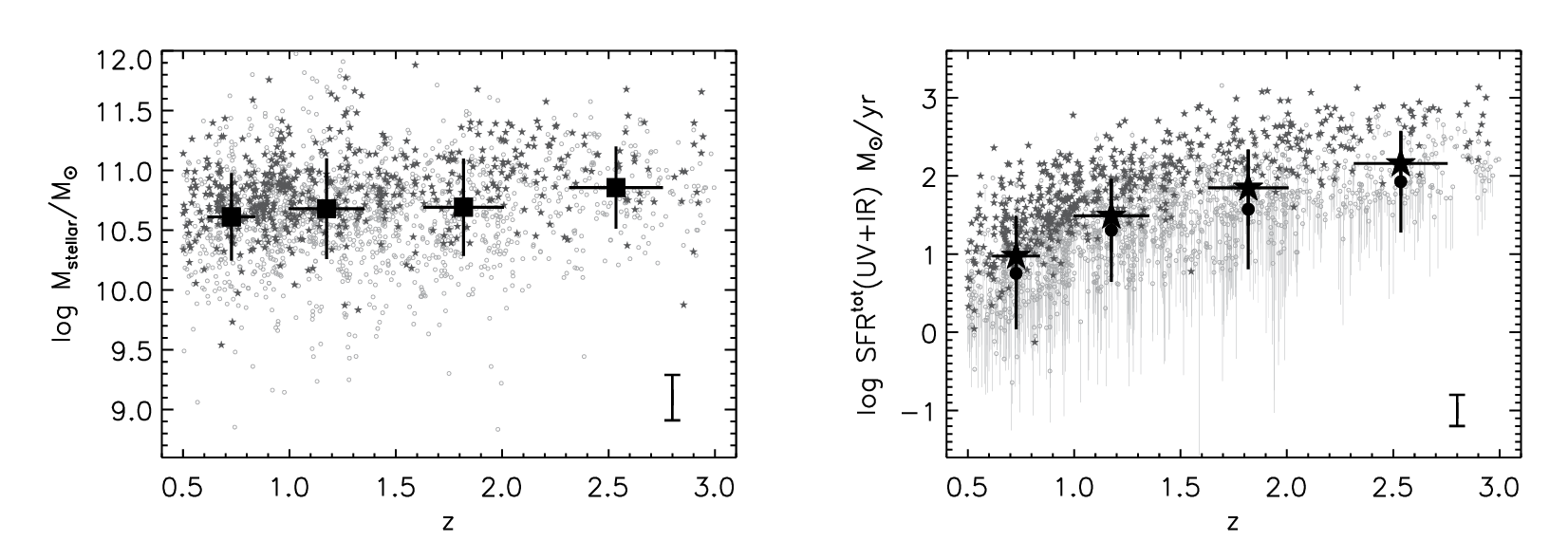}
\caption{Stellar mass and SFR distribution of our sample of Type 2 AGN host galaxies, normalized to the total area. The stellar mass distribution of our sample is shown in thick red histograms in top left panel. The distribution of all galaxies from the COSMOS catalog \citep{Laigle16} is also shown in gray shaded histogram for comparison. We also show the distribution of Type 2 AGNs in the COSMOS field from Bongiorno et al. (2012; blue) and Lusso et al. (2011; yellow). In top right panel, we show the SFR distributions, split into four redshift bins. In the bottom panels, the individual sources are indicated with filled stars ({\it Herschel}-detected) and open circles ({\it Herschel}-undetected) as a function of redshift. The range of SFRs for {\it Herschel}-undetected sources are indicated with gray error bars. Black stars indicate mean values of SFRs of {\it Herschel}-detected sources combined with possible maximum SFR of {\it Herschel}-undetected sources, while black circles represent that of SFRs of {\it Herschel}-detected sources combined with minimum SFR of {\it Herschel}-undetected sources. We also show the typical uncertainties in bottom right corner. 
}
\end{figure*}

In Figure 5, we show the stellar mass distribution (top left) and the total SFR distribution (top right) for our sample of Type 2 AGN host galaxies, normalized to the total area. For comparison, the stellar mass distributions of all galaxies in the COSMOS field \citep{Laigle16} are shown in gray shaded histogram. The distributions of Type 2 AGNs in the COSMOS field from \citet{Bongiorno12} and \citet{Lusso11} are also indicated with blue and yellow histograms, respectively. We also show the redshift evolution of stellar masses and SFRs in the bottom panels of Figure 5. Individual sources are indicated with gray filled stars ({\it Herschel}-detected; which are detected at least in one {\it Herschel} band) and circles ({\it Herschel}-undetected), and the range of SFRs for {\it Herschel}-undetected sources are also indicated with gray lines. Black stars represent the mean and the standard deviation of SFRs for the {\it Herschel}-detected sources combined with maximum SFRs of {\it Herschel}-undetected sources, indicating maximum mean SFRs. The minimum mean SFRs, of which the combination of SFR of the {\it Herschel}-detected source and minimum SFRs of {\it Herschel}-undetected source, are indicated with black circles. The typical uncertainties for the stellar masses ($\sim$0.19 dex) and the SFRs (for the {\it Herschel}-detected sources; $\sim$0.20 dex) are shown in the bottom right corner. The stellar mass of our sample ranges from $\sim10^{9}$ to $\sim10^{12}$ M$_{\odot}$, peaking at higher masses ($\sim5\times10^{10}$ M$_{\odot}$) compared to normal galaxies, consistent with results from \citet{Bongiorno12} and \citet{Lusso11}. There is a lack of significant evolution of stellar masses of Type 2 AGN host galaxies with redshift, which are relatively massive since $z\sim3$, indicating that they might have already experienced substantial growth at higher redshift ($z>3$). Our sample of Type 2 AGN host galaxies spans a wide range of SFRs, peaking at higher values toward higher redshifts. We note that the measurement of the SFR has considerably larger uncertainties than that of stellar mass, because it depends on the {\it Herschel} detections, SFRs could be inherently biased against higher values, while a significant fraction ($\sim$75\%) of our sample are faint in the far-IR photometry, which could have lower SFRs.

\section[]{The SFR-M$_{stellar}$ relation}

\begin{figure*}
\centering
\includegraphics[width=1\textwidth]{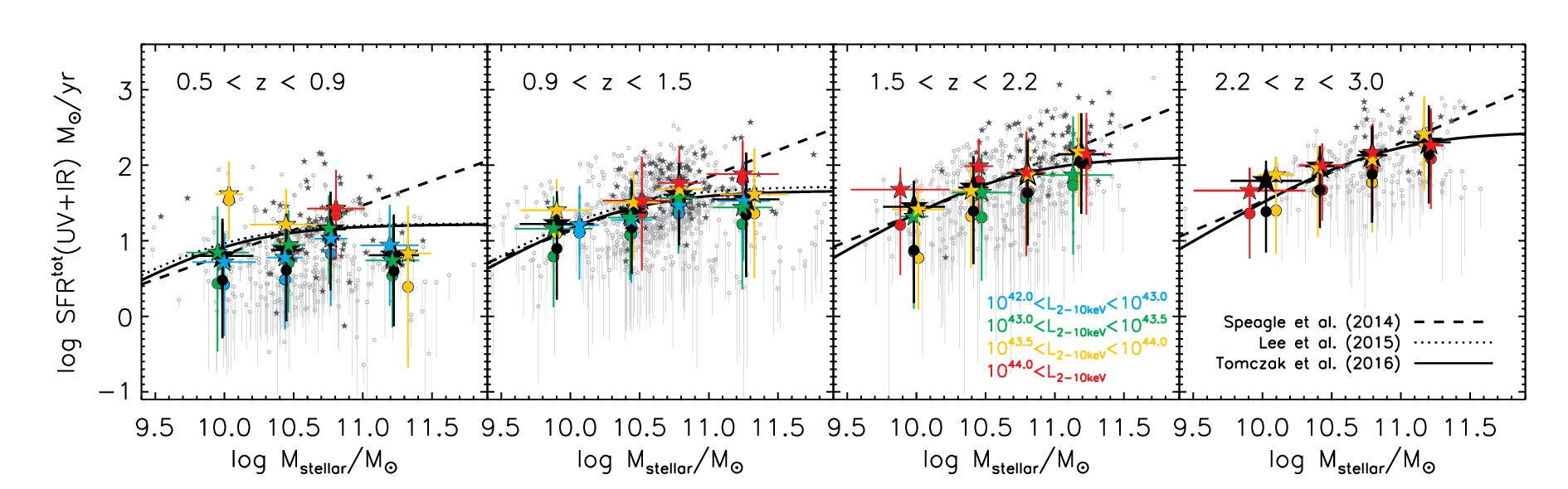}
\includegraphics[width=1\textwidth]{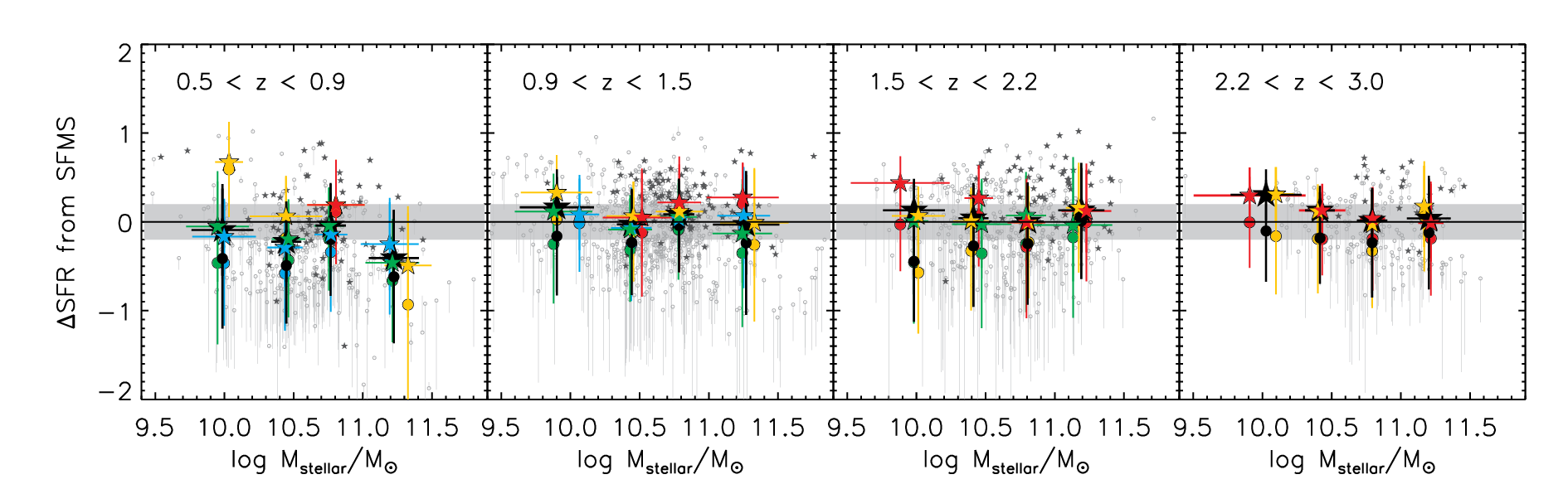}
\caption{{\it Top}: SFR versus stellar mass of our sample of Type 2 AGN host galaxies in the four redshift bins. Gray filled stars indicate the individual sources which are detected in the far-IR {\it Herschel} photometry, and gray open circles represent the possible maximum SFR for the sources which are not detected in any {\it Herschel} bands. The range of SFRs (i.e. from minimum to maximum) for {\it Herschel}-undetected sources are indicated with gray error bars. We indicate the star-forming MS relationships from Speagle et al. (2014; dashed line), Lee et al. (2015; dotted curve), and Tomczak et al. (2016; solid curve) for comparison. Black stars indicate mean values of SFRs of {\it Herschel}-detected sources combined with possible maximum SFR of {\it Herschel}-undetected sources, while black circles represent that of SFRs of {\it Herschel}-detected sources combined with minimum SFR of {\it Herschel}-undetected sources. Black thick error bars represent the range of mean SFRs which account for the maximum and minimum SFRs of the {\it Herschel}-undetected sources. We also display the mean SFRs for the sources at each X-ray luminosity bin (colored stars). {\it Bottom}: SFR offsets ($\Delta$SFR) relative to the star-forming MS of \citet{Tomczak16}. The gray shades mark the $\Delta$SFR$\sim\pm$0.2 dex. 
}
\end{figure*}

To investigate the effects of AGNs on the star formation in galaxies, we explore the distribution of our sample of Type 2 AGN host galaxies on the SFR-M$_{stellar}$ diagram compared to normal star-forming galaxies. Originally the star-forming MS studies concluded that the SFR increases with stellar mass as a single power law, while the log~SFR--log~M$_{stellar}$ slope and the normalization vary based on the redshifts, sample selection, choice of stellar initial mass function, and SFR indicators (for a summary see \citealt{Speagle14}). Recent studies have suggested that the SFR-M$_{stellar}$ relation flattens towards the high-mass end \citep{Whitaker14, Lee15, Tomczak16}. For example, \citet{Lee15} examine the star-forming MS, of which the total SFRs are determined by combination of the obscured SFRs using {\it Herschel} far-IR photometry and the unobscured SFRs from UV observations, using a large sample of $\sim$62,000 star-forming galaxies in the COSMOS field. The SFR indicator used in the \citet{Lee15} work is consistent with the one used for the CCLS sample. They find that the slope of the MS is dependent on stellar mass, such that it is steeper at low stellar masses and appears to flatten at stellar masses above $M_{stellar}\sim10^{10.3}M_{\odot}$, suggesting a curvature of the star-forming MS with a flat slope at the high mass end (see also \citealt{Whitaker14}). Furthermore, \citet{Tomczak16} present similar measurements of the star-forming MS up to $z\sim4$ using far-IR photometry from the {\it Spitzer} and {\it Herschel} observatories. They also suggest that the slope of star-forming MS becomes shallower above a turnover mass that is in the range from $10^{9.5}-10^{10.8} M_{\odot}$. 

We show SFRs and stellar masses of our sample of Type 2 AGN host galaxies, split into four redshift bins in the upper panels of Figure 6. The individual sources are indicated with filled gray stars when the sources are detected in {\it Herschel} far-IR photometry, while the circles represent the possible maximum SFR for the sources detected only up to 24 $\mu$m. The range of SFRs (i.e. from minimum to maximum) for {\it Herschel}-undetected sources is indicated with gray bars. We indicate the star-forming MS relationships from Tomczak et al. (2016; solid curve) and Speagle et al. (2014; dashed line) for comparison. The relation reported in the \citet{Lee15} study is also indicated with dotted curves at the low redshift bins. We show mean values of the combination of the SFR of {\it Herschel} detected sources (filled gray stars) and the maximum SFRs of the {\it Herschel}-undetected sources (open gray circles) in the stellar mass bins (black stars). Black circles mark the mean values of the combination of SFR of {\it Herschel} detected sources (filled gray stars) and the minimum SFRs of the {\it Herschel}-undetected sources. The black thick error bars represent the range of mean SFRs which account for the maximum and minimum SFRs for the {\it Herschel}-undetected sources. We also display the mean SFRs for the sources at each X-ray luminosity bins in the stellar mass bins (colored stars). 

In the lower panels of Figure 6, we show the SFR offset ($\Delta$SFR) for the AGN host galaxies relative to the star-forming MS of \citet{Tomczak16}. The gray shades mark the intrinsic scatter ($\sim$0.2 dex) of the star-forming MS. Most previous studies have found no clear evidence for a correlation between the X-ray luminosity and the SFR of the AGN host galaxy \citep{Lutz10, Shao10, Mullaney12, Rosario12, Rovilos12, Harrison12, Stanley15}. Our results indicate that there is no significant difference in the SFRs with respect to X-ray luminosity. Interestingly, it seems that there is a tendency for luminous ($10^{43.5}<L_{2-10keV} ({\rm erg/s}) <10^{44.0}$) AGN host galaxies to deviate from the star-forming MS relation in the range $0.5<z<0.9$. In this redshift range, AGN host galaxies with M$_{\rm stellar}/M_{\odot}<10^{10.5}$ show higher SFRs than star-forming MS galaxies, while massive AGN host galaxies (M$_{\rm stellar}/M_{\odot}>10^{11}$) seem to have SFRs that lie below the star-forming MS relation.

Type 2 AGN host galaxies, on average, seem to have SFRs that lie on the star-forming MS at all redshifts, consistent with previous studies (e.g. \citealt{Xue10, Mainieri11, Mullaney12, Rosario13}), but with much broader dispersions. \citet{Mullaney15} found that AGN host galaxies with ${\rm log~M_{stellar}/M_{\odot}\gtrsim10.3}$ show significantly broader SFR distributions compared to the star-forming MS galaxies, compared to normal galaxies (see also \citealt{Shimizu15}). We note, however, that Type 2 AGN host galaxies at high mass bins remain on the star-forming MS, when taking into account the dependence of the slope of the star-forming MS on stellar mass \citep{Whitaker14, Lee15, Tomczak16}. The selection effects and observational biases can be important since a significant fraction ($\sim$75\%) of our sample are not detected in far-IR photometry, which is crucial for precise measurements of the SFRs. The SFR distribution, therefore, is much broader than that of star-forming MS galaxies, when taking into account the fact that the SFRs of the {\it Herschel}-undetected sources could ultimately be much lower (i.e. minimum SFRs). Overall, Type 2 AGN host galaxies remain on the star-forming MS over a broad redshift range, indicating no sign of strong SFR enhancements in the redshift range $0.5<z<3.0$. 

\section{Discussion}

We discuss the star formation in Type 2 AGN host galaxies and the implications of the growth of black holes and galaxies over cosmic time. We show that the majority of Type 2 AGN host galaxies seem to reside along the star-forming MS, consistent with previous studies (e.g. \citealt{Mainieri11, Mullaney12, Rosario13}). While the ``flattening" in the star-forming MS at high masses could be interpreted as a consequence of quenching the star formation in massive galaxies (e.g. \citealt{Whitaker14, Lee15, Schreiber15}), the SFRs of Type 2 AGN host galaxies are consistent with those expected from normal star-forming galaxies in most stellar mass bins up to $z\sim3$, indicating no clear signature for enhanced or suppressed SFRs compared to normal star-forming galaxies. This can be interpreted by internal secular processes, which might be responsible for driving both star formation and nuclear activity in Type 2 AGN host galaxies. These results are consistent with the weak link between merger features and the modest AGN activity. From previous works in the literature (e.g. \citealt{Cisternas11, Mainieri11, Schawinski12, Fan14, Villforth14}), the majority of AGN host galaxies do not show significant merger features, indicating that mergers do not dominate the triggering of AGN activity, at least for moderate-luminosity AGNs. Allevato et al. (2011, 2012, 2016) further point out that moderate-luminosity AGNs inhabit group-sized halos ($10^{13-13.5}~{\rm M_{\odot}}$), almost independent of redshift up to $z\sim5$. This also implies that major mergers cannot be the main driver of the evolution of AGNs. 

However, this result could be also interpreted by the different timescales and the spatial scales associated with the star formation and nuclear activity \citep{Hickox14} in the sense that most AGN vary on a timescale much shorter ($\sim10^{5}$yr) than that of star formation ($\sim100$Myr) (e.g., \citealt{Hickox09, Aird12, Bongiorno12}). According to this scenario, all episodes of star formation and AGN activity could be intimately connected at any time. Furthermore, we should point out that these could be driven by the selection biases, mainly due to the interplay between the limited X-ray luminosity, Eddington ratio, SFRs and stellar masses of AGN host galaxies (e.g. \citealt{Lauer07, Xue10}). While AGNs preferentially reside in massive galaxies, when considering in the same stellar mass bins, SFRs of AGN host galaxies indicate no significant difference compared to normal star-forming galaxies. \citet{Xue10} also found that for mass-matched samples, the SFRs of AGN host galaxies are similar to those of non-AGN galaxies at $z\sim1-3$, consistent with our results. We further consider different X-ray luminosities to minimize potential luminosity-dependent effects. Within each stellar mass bin, we subdivide our sample into bins of the X-ray luminosity. With luminosity-selection effects taken into account, we find that there is no clear signature for a correlation between the AGN luminosity and the SFRs of AGN host galaxies. The Eddington ratio could be a factor that creates a bias against low-luminosity AGNs accreting at the lowest Eddington ratios at high redshift. Our AGN sample may also bias against the heavily obscured, Compton-thick sources, which might be missed by X-ray selection (e.g. \citealt{Treister04, Kocevski15}). However, at least for our sample of moderate-luminosity X-ray selected AGNs, we find that there is no significant difference between AGN hosts and normal star-forming galaxies.

From the perspective of our investigation on the star-formation in Type 2 AGN host galaxies, we propose that the relatively massive galaxies have already experienced substantial growth by major mergers, which are capable of triggering both a significant starburst and high accretion AGN activity at higher redshift ($z>3$), and grow slowly through secular fueling processes hosting moderate luminosity AGNs. \citet{Aird12} present that AGN Eddington ratios are independent of stellar masses of their hosts at $z<1$, suggesting that the same physical processes regulate AGN activity in galaxies at stellar masses ${\rm 9.5\lesssim log~M_{stellar}/M_{\odot}\lesssim12.0}$. \citet{Suh15} further point out that a substantial fraction of massive black holes accreting significantly below the Eddington limit at $z<2$, suggesting that modest AGN activity can be triggered via internal, secular processes in massive galaxies. This is also compatible with the lack of significant evolution of stellar masses of Type 2 AGN host galaxies. Our results suggest that the majority of Type 2 AGN host galaxies at $z<3$ might be driven more by internal secular processes, implying that they have substantially grown at much earlier epoch.

\section{Summary}

We present the host galaxy properties of a large sample of $\sim$2300 X-ray-selected Type 2 AGNs out to $z\sim3$ in the CCLS in order to examine whether AGN activity can significantly enhance or quench star formation in galaxies. To derive the physical properties of AGN host galaxies, we develop a multi-component SED fitting technique to disentangle the nuclear emission from the stellar light, and derive host galaxy properties. Specifically, we use multi-band photometry (from near-UV through the far-IR) to decompose the entire SED into separate components with nuclear AGN emission, the host galaxy's stellar populations, and a starburst contribution in the far-IR. We derive stellar masses of our sample in the range $9 < \log M_{\rm stellar}/M_{\odot} < 12$ with uncertainties of $\sim$0.19 dex. The SFR is estimated by combining the contributions from UV and IR luminosity. Our sample of Type 2 AGN host galaxies span a wide range of SFRs ($-1 < \log {\rm SFR}~(M_{\odot}/{\rm yr}) < 3$) with uncertainties of $\sim$0.20 dex.

We explore the distribution of AGN host galaxies on the SFR-stellar mass diagram compared to the normal star-forming galaxies. Overall, Type 2 AGN host galaxies seem to have SFRs that lie on the star-forming MS up to $z\sim3$, independent of X-ray luminosities. Our results indicate that AGN host galaxies do not show clear signature for enhanced or suppressed SFRs compared to normal star-forming galaxies. 


\acknowledgments

This work was supported in part by NASA {\it Chandra} grant number GO3-14150C, GO3-14150B, and also GO5-16150A. K. S. acknowledges support from Swiss National Science Foundation Grants PP00P2\_138979 and PP00P2\_166159. E.L. is supported by a European Union COFUND/Durham Junior Research Fellowship (under EU grant agreement no. 609412). 


\begin{turnpage}
\begin{deluxetable*}{lrrccccc}
\tabletypesize{\scriptsize}
\tablewidth{0pt}
\tablecaption{Type 2 AGN host galaxy properties derived from the SED fitting}
\tablehead{
    \colhead{\textbf{ID}} & 
    \colhead{\textbf{log M$_{\rm stellar}$ (M$_{\odot}$)}} &
    \colhead{\textbf{log SFR$^{\rm tot}$ (M$_{\odot}$ yr$^{-1}$)}} &
    \colhead{\textbf{log L$_{2-10~keV}$ (erg s$^{-1}$)}} &    
    \colhead{\textbf{log L$_{2300}$ (erg s$^{-1}$)}} &
    \colhead{\textbf{log L$_{5100}$ (erg s$^{-1}$)}} &
    \colhead{\textbf{log L$_{6\mu m}$ (erg s$^{-1}$)}} &
    \colhead{\textbf{log L$_{IR}$ (erg s$^{-1}$)}}}
\startdata
  CID-3  &  9.98$^{+0.041} _{-0.10}$ & 0.48$^{+0.06} _{-0.29}$ & 42.95 & 43.50 & 43.75 &  43.17 & 43.82 \\
  CID-12 & 10.44$^{+0.11} _{-0.03}$ & 1.29$^{+0.00} _{-0.43}$ & 43.07 & 44.14 & 44.04 & 44.06 & 44.69  \\
  CID-22 & 10.36$^{+0.05} _{-0.08}$ & 1.07$^{+0.01} _{-0.30}$ & 43.76 & 44.04 & 44.12 & 43.80 & 44.44  \\
    ... & ... & ... & ... & ... & ... & ...   \\
 \enddata
  \tablecomments{Parameters derived from the SED fitting. The columns are: (1) {\it Chandra} source ID from \citet{Civano16}; (2) host galaxy stellar mass; (3) host galaxy SFR; (4) absorption-corrected X-ray luminosity in 2--10 keV band; (5) rest-frame UV luminosity at 2300\AA of the host galaxy component (6) rest-frame luminosity at 5100\AA of the host galaxy component (7) rest-frame 6$\mu$m luminosity of the AGN component (8) IR luminosity, L$_{8-1000\mu m}$, of the star-burst component.}
\end{deluxetable*}
\end{turnpage}


\end{document}